\documentstyle[11pt,twoside,pasp3D,epsf]{article}

\begin{document}

\title{LISA A low Resolution MIR/FIR Imaging Spectrograph for SOFIA}

\author{A. Krabbe and J. Wolf}
\affil{German Aerospace Center, Institute of Space Sensor Technology
and Planetary Exploration, Rutherfordstra$\ss$e 2,12489 Berlin -
Adlershof, Germany}

\begin{abstract}
LISA is a proposed low resolution (R$\sim$15-30) imaging spectrometer for
SOFIA, the American-German Stratospheric Observatory for Infrared Astronomy.
Covering the wavelength range from 40$\micron$ to 220$\micron$ with three
channels, LISA provides diffraction limited spectroscopy for each pixel
within a rectangular field of view (FOV). The FOV sizes for the channels
are 18$\times$11$\arcsec$ for 40 to 70$\micron$, 32$\times$19$\arcsec$ for
70-120$\micron$, and 22$\times$22$\arcsec$ for 120-220$\micron$, each
channel upgradable to about twice the linear size.

LISA will be able to address a variety of astrophysical topic,
in particular on faint targets. Spatially resolved temperature distribution of
dust emission can be studied in the environments of young and evolved stars,
star forming regions, as well as in nuclear regions of nearby galaxies.
Spectral energy distributions (SEDs) of distant galaxies will be determined in
search for cold gas and dust components. LISA$^{\prime}$s spectral
resolution will be high enough
to search for broad spectral features (e.g. PAHs, ices). The optical design,
the mechanical layout, the estimated sensitivity, and the scientific potential
of such an instrument are briefly discussed.
\end{abstract}

\section{Introduction}
SOFIA, the $\underline{S}$tratospheric $\underline{O}$bservatory
$\underline{F}$or $\underline{I}$nfrared $\underline{A}$stronomy, will be the
biggest and probably the only platform with regular access to the entire
MIR and FIR wavelength range between 25 and 250$\micron$ in both celestial
hemispheres for the next two decades. From 2002 on it will open a new era
in MIR/FIR astronomy, providing the highest available spatial resolution
in combination with excellent sensitivity. SOFIA is a 2.7m telescope
onboard a Boeing 747SP flying at an altitude of between 31000 up to 45000
feet during typically 7 hours observing flights. 10 facility, PI
and special purpose instruments have already been selected and approved
for first light.

Here we present a concept study of LISA, a highly efficient
$\underline{L}$ow resolution $\underline{I}$ntegral field
$\underline{S}$pectrometer for SPIC$\underline{A}$,
which is considered to become part of the proposed Spectral-Photometric
Far-Infrared Camera SPICA (see Wolf et al., this issue) for SOFIA. LISA
will be a
self-contained subunit of SPICA, complementing it with diffraction
limited integral field spectroscopic capabilities at low spectral
resolution of about R$\sim$20.

\section{Instrument Concept}
The wavelength range covering 2.5 octaves is divided into 3 separate
intervals, each of which is fed into a separate grating spectrometer unit.
The spectral resolution in each channel varies between 16 and 28 thus covering
the whole spectral range with about 36 independent spectral resolution
elements. Fully sampled spectra can be obtained by tilting the grating
slightly by a fixed amount, such that the spectra are shifted by half a
pixel across the array in the dispersion direction. The current design is
optimized for 32$\times$32 pixel Ge:Ga detectors with 350$\micron$ pixel pitch
and 16$\times$16 stressed Ge:Ga with 4mm pitch, but can easily be
extended to accommodate larger e.g. 64$\times$64 pixel detectors. The only
moving part within LISA will be the grating tilt mechanisms.

\begin{figure}
\plotone{Krabbe.Fig1.EPSF}
\caption{LISA$^{\prime}$s 70 - 120$\micron$ spectrometer channel. }
\end{figure}

An imaging spectrometer was given preference over a set of narrow band
filters for several reasons:
\begin{itemize}
\item Flat-top high transmission narrow-band filters with
$\lambda/\Delta\lambda \sim 6 - 20$ covering the spectral range
40$\micron$ - 220$\micron$ are not available. The 3 broad-band
($\lambda/\Delta\lambda \sim$2) order sorting filter for LISA will have
higher transmission and a better blocking than existing narrow-band
filter.

\item The data sets obtained with an imaging spectrometer are more
consistent and always complete. Since this instrument will obtain a data
cube during each individual integration, varying observing conditions will
effect all data consistently in the same fashion. Such changes during the
flight include variation of seeing, air mass, water vapor \& atmospheric
transmission, residuals in telescope tracking, pointing, image rotation \&
structural bending, as well as detector performance \& temperature variations.
Under unfavorable conditions, these modulations will introduce systematic
noise into nonmultiplexed data, thus lowering the effective signal/noise
ratio significantly.

\item The integration time per target is limited to typically 3 hours per
target and the risk of unexpected technical events on an airborne telescope
is slightly higher compared to a ground based observatory. Under such
conditions
most efficient and multiplexing observing techniques are strongly
advised. In addition, spectral multiplexing avoids overheads created by the
need to
frequently changing the setting of the instrument. Under less than optimal
observing conditions, consistent spectral information is at least as important
as enhanced spatial resolution.

\item Most targets under consideration only cover a fraction of the detector
area, in particular if even larger arrays become available. Imaging
spectroscopy uses the whole array since it distributes spectral as well
as spatial information across it and thus maximizes the observing
efficiency at any time.

\item An imaging spectrometer does not require an excellent pointing of
the telescope to put a target on a narrow slit. Its field of view (FOV)
is much wider and the pointing effort is more comparable to an instrument
using narrow band filters.

\item The imaging capabilities of such a spectrometer can be used to acquire
the target or to check its position during the integration. A two dimensional
image can always be extracted by collapsing the spectral information and
rearranging the pixels according to their proper location on the sky.
\end{itemize}

%

\begin{figure}
\plotone{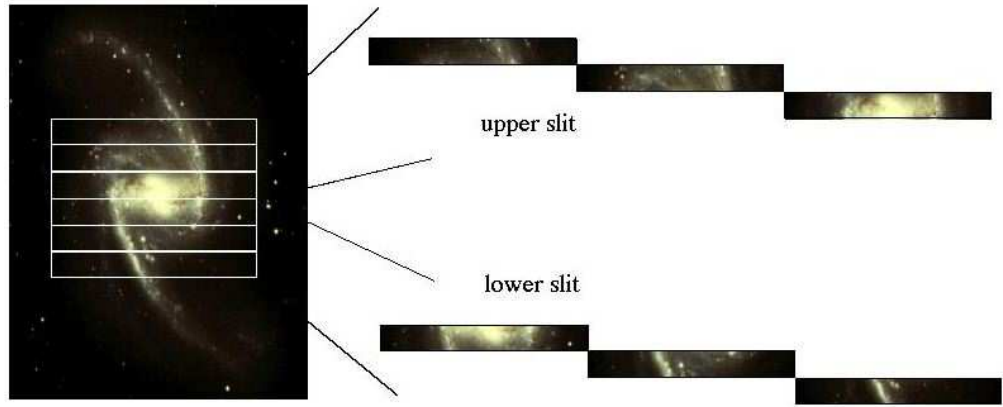}
\caption{The working principle of the image slicer. A rectangular field in the
sky is split into six individual stripes, which are realigned to form two
parallel
entrance slits at the input of a grating spectrometer.}
\end{figure}

\section{Optical design}
The basic design of LISA is that of a long slit spectrometer (Figure 1)
with two parallel input slits. The spectrometer is designed such that both
entrance slits are dispersed into nonoverlapping spectra on the array. A
system of optical flat mirrors, the image slicer, rearranges the two slits
into one rectangular FOV on the sky. Figure 2 explains the slicing technique
on the slicer for the 70 - 120$\micron$ wavelength range. Figure 3 shows a
close up view of the image slicer itself. For a 32$\times$32 Ge:Ga detector
it will be composed of two sets of 6 gold coated plane mirrors and a
mechanical
support structure all made of diamond turned aluminum. The first set of
mirrors, located in an intermediate focus of the telescope, slices a
rectangular image into 2$\times$3 strips and deflects them into different
directions while the second set of mirrors picks up light from the individual
strips and aligns them into two contiguous long strips such that the pupils
of the 6 single strips are coincident. The distribution of the spectra
across the detector is illustrated in Figure 4.

Since the optical train is telecentric at the location of the entrance slit
(see Wolf et al. 1998), the recombined 6 single strips are optically
indistinguishable from two single long slits. The fact that the resulting
slits look stair-like shifts the spectra on the detector
by only a small amount with respect to each other. Each single mirror
of the first set is about 12 mm long and 1.2 mm high. The area centers of
the second set of mirrors all lie on a rotating parabola such that the
focus of the parabola coincides with the telescope focus and the position
of the entrance pupil of the unsliced image is retained at -$\infty$.
The FOV of LISA was chosen 10$\times$6 pixels. Such a rectangular format is
more flexible since most of the targets do not show circular symmetry. A
quadratic FOV covering 8$\times$8 pixels is another possible configuration
which can be realized with 2$\times$4 slitlets if each slitlet is 8 pixels
long.

\begin{figure}
\begin{center}
{\centering \leavevmode\epsfxsize=10cm \epsfbox{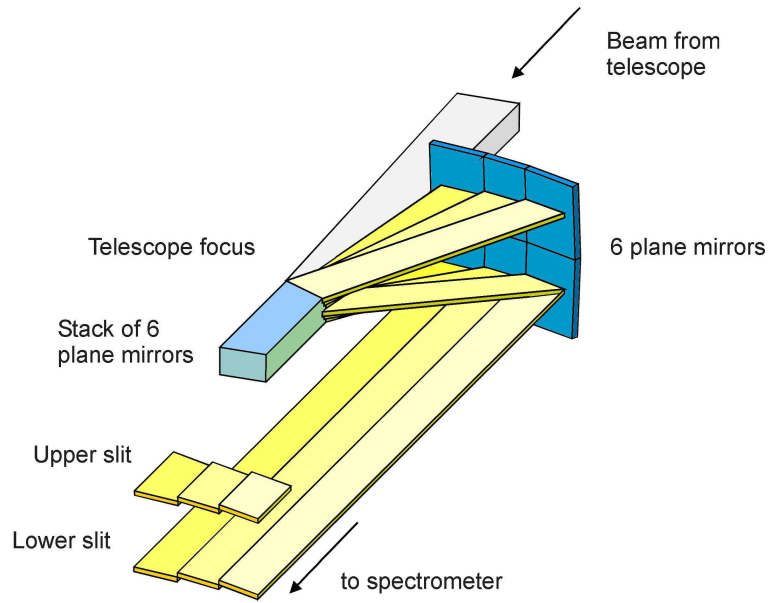}}
\end{center}
\caption{Close-up view of the image slicer. 12 plane mirrors rearrange the
image information. Compare with Figure 2.}
\end{figure}

\begin{table}
\caption{Current basic parameters of LISA} \label{tbl-1}
\begin{center}\scriptsize
\begin{tabular}{crrrrrrrrrrr}
Range & 40-70$\micron$ &  & 70-120$\micron$ &  & 120-220$\micron$ \\
\tableline
Pixel scale $[\arcsec]$ &1.8 &  &3.2  &  &5.6 &  \\
$\Delta\lambda [\micron]$ & 2.3 &  & 4.0 &  & 6.5 &  \\
Detector &Ge:Ga &   &Ge:Ga  &  & sGe:Ga  &  \\
Detector matrix [pixel$^2$] & 32$\times$32 & 64$\times$64 & 32$\times$32 &
64$\times$64 & 16$\times$16 & 32$\times$32\\
Pitch $[\micron]$ & 350 &  & 350 &  &4.0  &  \\
Scale at entrance slit $[\arcsec/mm]$ & 2.8 &  & 2.8  &  & 2.8 &  \\
Field of view [pixel] & 10$\times$6 & 21$\times$12 & 10$\times$6 &
21$\times$12 & 4$\times$4 &10$\times$6 \\
\hspace{1cm} $[\arcsec]$ & 18$\times$11 & 38$\times$20 & 32$\times$19 &
38$\times$20 & 22$\times$22 & 56$\times$34 \\
Slitlet height [mm] & 0.7 &  & 1.2 & 2.0 &  \\
Slitlet length [mm] & 7 & 14 & 12 & 24 & 8 & 20 \\
f$_{collimator}$ [mm] & 360 &  & 360 &  & 360 &  \\
f$_{camera}$ [mm] & 193 &  & 108 &  & 180 &  \\
\end{tabular}
\end{center}
\end{table}

\begin{figure}
\begin{center}
{\centering \leavevmode\epsfxsize=10cm \epsfbox{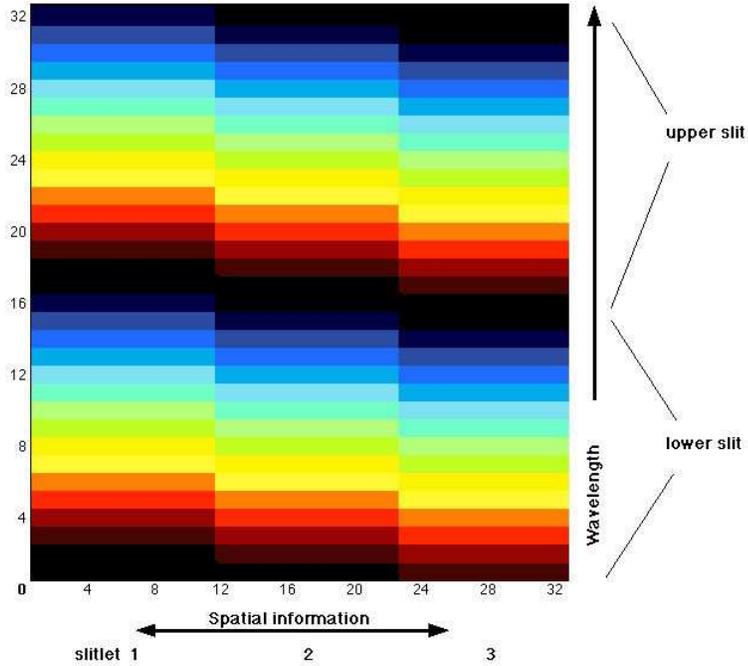}}
\end{center}
\caption{Distribution of the spectra across the focal plane of LISA on a
32$\times$32 pixel detector. The two entrance slits are dispersed on the same
detector but do not overlap. The stair-like geometry of each entrance slit
is reflected in the spectral offsets between individual slitlets.}
\end{figure}

The current basic parameters of the 3 channels are given in Table 1 for two
different detector sizes. The spectrometer itself will be a modified
Cerny-Turner
type with the grating in off-axis Littrow configuration. The tilted exit
slit can be accommodated for by rotating the slicer with respect to the
optical axis by the off-axis angle (not shown in Figure 1). Collimator and
camera mirror are off-axis parabolas. Diffraction effects at the slicer have
to be taken into account since in the worst case the height of a slitlet is
only about 10 wavelengths across. Behind the slit the beam will flare to f/11
instead of f/30 perpendicular to the slit. This leads to an elliptical pupil
on the grating with an axis ratio of about 2.5:1. The background on the
detector,
however, will not be enhanced since the cold Lyot-stop further up in the
beam has already limited the background flux.

\section{Scientific potential}

The current design of LISA is based on the experience with 3D, a near-infrared
imaging spectrometer (Krabbe et al. 1996, Weitzel et al. 1996). The expected
optical transmission of the slicer will be about 95\%, the area covering
factor of the detector will be $\>$80\% yielding an optical efficiency of
$\>$75\%. If the target fits within LISA's FOV, the multiplex gain over a set
of filters is about 13. Between 12 and 14 spectral positions are recorded
per integration under background limited conditions. An instrument
using $n$ narrow band filters across the spectral range 40 -
220$\micron$ will be about $n$/3 times less efficient provided that the other
losses roughly balance each other. With larger arrays the multiplex advantage
of an imaging spectrometer will be even higher by about the ratio of the
number
of pixels.

\begin{figure}
\begin{center}
{\centering \leavevmode\epsfxsize=10cm \epsfbox{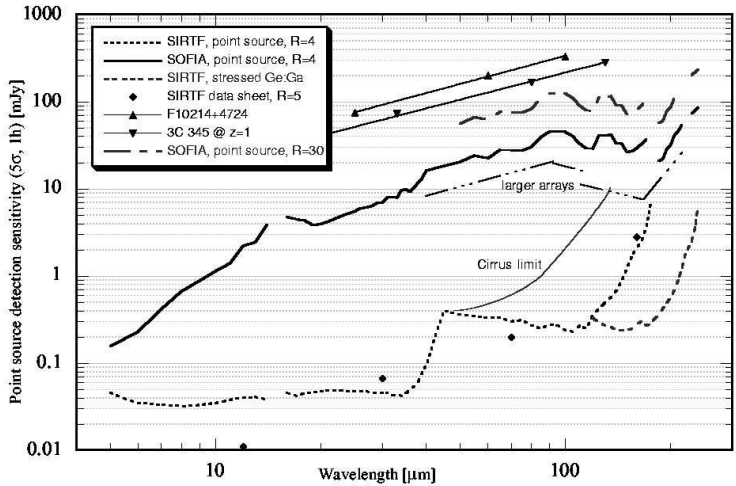}}
\end{center}
\caption{SOFIA$^{\prime}$s point source sensitivity for broad band imaging and
low-resolution spectroscopy. Curves for the SIRTF mission are given for
comparison.
The triple-dot-dashed curve denotes the increase in sensitivity of SOFIA for
surveys due to spatial multiplexing for 64$\times$64 pixel Ge:Ga and a
32$\times$32
stressed Ge:Ga array. The "Cirrus limit" curve is a rough estimate (see e.g.
Franceschini et al. 1991 and Herbstmeier et al. 1998 for details). The far
infrared sensitivity for a
spectral resolution of 30 is denoted by the dot-dashed curve.}
\end{figure}

A model calculation on SOFIA$^{\prime}$s point source sensitivity using
conservative assumptions is shown in Figure 5.  For a spectral resolution of 16
instead of 4 the sensitivity curves are for 2.5$\sigma$ instead of
5$\sigma$ and
1hour integration.  The broad band SEDs of IRAS10214+4724 and 3C-234 redshifted
to z=1 are shown to illustrate SOFIA$^{\prime}$s sensitivity.  All IRAS sources
lie well above these curves.  Between 40$\micron$ and 100$\micron$,
LISA$^{\prime}$s sensitivity ranges between 40mJy and 70mJy per beam.

LISA will be able to determine the spectral energy distribution (SED) as
well as
broad spectral features in galactic and extragalactic targets.  On extended
objects, the spectral images can be decomposed into components of different
temperature and spatial morphology.  Applications are widespread.  In star
forming regions, the interaction between young stars and their environments,
their outflows and dynamical evolution (e.g. TTauri stars, HH objects, AeBe
stars) can be investigated in detail.  Dust disks and the evolution of
protoplanetary conditions can be studies on Vega like main sequence stars (e.g.
Vega, $\beta$Pic).  Not much is known about the expanding envelopes of more
massive evolved stars during their AGB phase and evolution into protoplanetary
nebulae.  Interaction between ejecta from violent supernovae explosions and
interstellar matter leads to progressing shockfronts, which heat up and process
dust and gas of the surrounding molecular clouds.  In our solar system low
resolution spectroscopy will be an important tool in analyzing the composition
of solids and ices on planetary surfaces and in their atmospheres as well as on
asteroids and in comets.

The dust content and temperature composition of external active and luminous
galaxies like Seyferts, starbursts,
or interacting systems will be investigated by obtaining the (spatially
resolved) SEDs of the
different components and thus constraining the physical processes involved in
the energy production. A special population are  cool and/or very red objects
(e.g. VERO's). Emphasis will be on identifying and studying the colder
components in these targets. LISA will thus complement the camera with an
efficient tool for analyzing the broad spectral characteristics of targets,
which will be discovered and/or studied with SPICA.

\end{document}